\documentclass{article}

\usepackage[top=3cm,bottom=3cm,left=3cm,right=3cm]{geometry}
\usepackage[characters]{ltl}
\usepackage{listings}
\usepackage{verbatim}
\usepackage{url}
\usepackage{multirow}
\usepackage[final]{hyperref}

\colorlet{codecolor}{blue!70!black}

\newcommand{\bools}{\ensuremath{\mathbb{B}}}
\newcommand{\nats}{\ensuremath{\mathbb{N}}}
\newcommand{\busses}{\ensuremath{\mathbb{U}}}
\newcommand{\temporals}{\ensuremath{\mathbb{T}}}
\newcommand{\signals}{\ensuremath{\mathbb{S}}}
\newcommand{\pats}{\ensuremath{\mathbb{P}}}
\newcommand{\arbitrary}{\ensuremath{\mathbb{X}}}
\newcommand{\signalids}{\ensuremath{\Gamma_{\signals}}}
\newcommand{\busids}{\ensuremath{\Gamma_{\busses}}}
\newcommand{\natids}{\ensuremath{\Gamma_{\nats}}}
\newcommand{\boolids}{\ensuremath{\Gamma_{\bools}}}
\newcommand{\temporalids}{\ensuremath{\Gamma_{\temporals}}}
\newcommand{\atypeids}{\ensuremath{\Gamma_{\mathcal{S}_{\arbitrary}}}}
\newcommand{\lstilde}{{\ensuremath{\color{codecolor} \raisebox{-9pt}{\scalebox{1.7}{\textasciitilde}}}}}
\newcommand{\tlsfsec}[1]{\ensuremath{\langle \text{\textit{#1}} \rangle}}
\newcommand{\tlsfid}[1]{\ensuremath{\langle \text{\textit{#1}} \rangle}}
\newcommand{\inputs}{\ensuremath{\mathcal{I}}}
\newcommand{\outputs}{\ensuremath{\mathcal{O}}}
\newcommand{\sep}{\ensuremath{\ \ | \ \ }}

\renewcommand{\subsubsection}[1]{\medskip \noindent {\bf #1.}}

\newcommand{\src}[1]{\texttt{\lstinline!#1!}}
\newcommand{\secref}[1]{Sect.~\ref{#1}}

\lstdefinestyle{TLSF}{
  belowcaptionskip=1\baselineskip,
  breaklines=true,
  frame=L,
  xleftmargin=\parindent,
  language=C,
  showstringspaces=false,
  basicstyle=\ttfamily\color{codecolor},
  keywordstyle=\ttfamily\color{codecolor},
  commentstyle=\ttfamily\color{codecolor},
  identifierstyle=\text\ttfamily\color{codecolor},
  stringstyle=\ttfamily\color{codecolor},
}

\begin{document}

\begin{center}
  \begin{huge}
    A High-Level LTL Synthesis Format: \\[0.5em]
    TLSF v1.0
  \end{huge}

  \vspace{2em}

  \begin{large}
    Swen Jacobs, Felix Klein
  \end{large}

  \vspace{0.9em}
  
  \verb!{jacobs,klein}@react.uni-saarland.de!

%
\end{center}

\vspace{0.2em}

\begin{abstract}
  We present the Temporal Logic Synthesis Format (TLSF), 
	a high-level format to describe synthesis problems via
  Linear Temporal Logic (LTL). The format builds upon standard LTL,
  but additionally allows to use high level constructs, such as sets
  and functions, to provide a compact and human readable
  representation. Furthermore, the format allows to identify
  parameters of a specification such that a single description can be
  used to define a family of problems. We also present a tool to
  automatically translate the format into plain LTL, which then can be
  used for synthesis by a solver. The tool also allows to adjust
  parameters of the specification and to apply standard
  transformations on the resulting formula.
\end{abstract}

\section{Overview}
We present the basic version of the Temporal Logic Synthesis Format (TLSF)
in \secref{sec:basicformat}. In
\secref{sec:semantics} we discuss the intended semantics of a
specification, defined in terms of different implementation models.
The full format is introduced in
\secref{sec:format}. The full format can be compiled automatically to
the basic format. Thus, while it is more convenient to write
specifications in the full format, for a synthesis tool it is
sufficient to support the basic format. We illustrate the main
features of the format in an example in \secref{sec:example}.  In
\secref{sec:tool}, we give an overview of the SyFCo Tool, which can
interpret the specification, possibly with respect to given parameter
values, and transform it to the basic format, as well as a number of
existing specification formats for backwards compatibility. Finally,
we discuss possible extensions of the format in
\secref{sec:extensions}.

\section{The Basic Format}
\label{sec:basicformat}

A specification in the basic format consists of an INFO section and a
MAIN section:
\begin{equation*}
  \tlsfsec{info} \tlsfsec{main} 
\end{equation*}

\subsection{The INFO Section}
\label{sec:basicinfo}

The INFO section contains the meta data of the specification, like a
title and some description\footnote{We use colored verbatim font to
  identify the syntactic elements of the specification.}. Furthermore,
it defines the underlying semantics of the specification (Mealy or
Moore / standard or strict implication) and the target model of the
synthesized implementation. Detailed information about supported
semantics and targets can be found in \secref{sec:semantics}.
Finally, a comma separated list of tags can be specified to identify
features of the specification, e.g., the restriction to a specific
fragment of LTL. A \tlsfid{tag} can be any string literal and is not
restricted to any predefined keywords.

\goodbreak

\vspace{1em}

\noindent
\lstinline!  INFO {!\\%
\lstinline!    TITLE:       "!$ \tlsfid{some title} $\lstinline!"!\\%
\lstinline!    DESCRIPTION: "!$ \tlsfid{some description} $\lstinline!"!\\%
\lstinline!    SEMANTICS:   !$ \tlsfid{semantics} $\\%
\lstinline!    TARGET:      !$ \tlsfid{target} $\\%
\lstinline!    TAGS:        !$ \tlsfid{tag} $%
\lstinline!,! $ \tlsfid{tag} $\lstinline!,!$ \ \ldots $\\%
\lstinline!  }!

\subsection{The MAIN Section}

The specification is completed by the MAIN section, which contains the
partitioning of input and output signals, as well as the main
specification, separated into environment assumptions, system invariants
and system guarantees. Multiple declarations and expressions need
to be separated by a '\lstinline!;!'.

\vspace{1em}

\noindent
\lstinline!  MAIN {!\\%
\lstinline!    INPUTS      { !%
$ ( \tlsfsec{boolean signal declaration} $\lstinline!;!$ )^{*} $\lstinline! }!\\%
\lstinline!    OUTPUTS     { !%
$ ( \tlsfsec{boolean signal declaration} $\lstinline!;!$ )^{*} $\lstinline! }!\\%
\lstinline!    ASSUMPTIONS { !%
$ ( \tlsfsec{basic LTL expression} $\lstinline!;!$ )^{*} $\lstinline! }!\\%
\lstinline!    INVARIANTS  { !%
$ ( \tlsfsec{basic LTL expression} $\lstinline!;!$ )^{*} $\lstinline! }!\\%
\lstinline!    GUARANTEES  { !%
$ ( \tlsfsec{basic LTL expression} $\lstinline!;!$ )^{*} $\lstinline! }!\\%
\lstinline!  }!

\vspace{1em}

\noindent
The ASSUMPTIONS, INVARIANTS and GUARANTEES subsections are optional.

\subsection{Basic Expressions}

A basic expression~$ e $ is either a boolean signal or a basic LTL
expression. Each basic expression has a corresponding type that is
$ \signals $ for boolean signals and $ \temporals $ for LTL
expressions.  Basic expressions can be composed to larger expressions
using operators.  An overview over the different types of expressions
and operators is given below.

\subsubsection{Boolean Signal Declarations}
A signal identifier is represented by a string consisting of lowercase
and uppercase letters~\mbox{('\lstinline!a!'-'\lstinline!z!'},
\mbox{'\lstinline!A!'-'\lstinline!Z!')},
numbers~\mbox{('\lstinline!0!'-'\lstinline!9!')},
underscores~\mbox{('\lstinline!_!')}, primes~\mbox{('\lstinline!'!')},
and at-signs~\mbox{('\lstinline!@!')} and does not start with a number
or a prime. Additionally, keywords like \lstinline!X!, \lstinline!G!
or \lstinline!U!, as defined in the rest of this document, are
forbidden.
An identifier is declared as either an input or an output signal. We
denote the set of declared input signals as $ \inputs $ and the set of
declared output signals as $ \outputs $, where
$ \inputs \cap \outputs = \emptyset $.  Then, a boolean signal
declaration simply consists of a signal identifier \tlsfid{name} from
$ \inputs \cup \outputs $.

\subsubsection{Basic LTL Expressions}
A basic LTL expression conforms to the following grammar, including
truth values, signals, boolean operators and temporal operators. For
easy parsing of the basic format, we require fully parenthesized
expressions, as expressed by the first of the following lines:
\begin{eqnarray*}
  \varphi & \equiv & 
  \text{\lstinline|(|} \varphi' \text{\lstinline|)|} \\
  \varphi' & \equiv & 
  \text{\lstinline|true|} \sep
  \text{\lstinline!false!} \sep
  s \text{\ \ \ for } s \in \inputs \cup \outputs \sep 
  \\ & &
  \text{\lstinline|!|} \varphi \sep
  \varphi \ \text{\lstinline!&&!} \ \varphi \sep
  \varphi \ \text{\lstinline!||!} \ \varphi \sep
  \varphi \ \text{\lstinline!->!} \ \varphi \sep
  \varphi \ \text{\lstinline!<->!} \ \varphi 
  \\ & & 
  \text{\lstinline!X!} \ \varphi \sep
  \text{\lstinline!G!} \ \varphi \sep
  \text{\lstinline!F!} \ \varphi \sep
  \varphi \ \text{\lstinline!U!} \ \varphi \sep
  \varphi \ \text{\lstinline!R!} \ \varphi \sep
  \varphi \ \text{\lstinline!W!} \ \varphi
\end{eqnarray*}
Thus, a basic LTL expression is based on the expressions 
true, false, and signals, composed with boolean operators (negation,
conjunction, disjunction, implication, equivalence) and temporal
operators (next, globally, eventually, until, release, weak until).
The semantics are defined in the usual way.  A formal definition of
the semantics of the temporal operators can be found in
Appendix~\ref{apx_ltl}.

\newpage

\section{Targets and Semantics}
\label{sec:semantics}

The TARGET of the specification defines the implementation model that
a solution should adhere to. Currently supported targets are Mealy
automata \mbox{(\lstinline!Mealy!)}, whose output depends on the
current state and input, and Moore automata
\mbox{(\lstinline!Moore!)}, whose output only depends on the current
state. The differentiation is necessary since realizability of a 
specification depends on the target system model. For example, every 
specification that is realizable
under Moore semantics is also realizable under Mealy semantics, but
not vice versa. A formal description
of both automata models can be found in Appendix~\ref{apx_memo}.

The SEMANTICS of the specification defines how the formula
was intended to be evaluated, which also depends on the target
implementation model.  We currently support four different semantics:
standard Mealy semantics \mbox{(\lstinline!Mealy!)}, standard Moore
semantics \mbox{(\lstinline!Moore!)}, strict Mealy semantics
\mbox{(\lstinline!Mealy,Strict!)}, and strict Moore semantics
\mbox{(\lstinline!Moore,Strict!)}. 

\subsection{Basic cases}

\subsubsection{Standard semantics}
If the semantics is (non-strict) \mbox{\lstinline!Mealy!} or 
\mbox{\lstinline!Moore!}, and the TARGET coincides with the semantics system 
model, then the specification is simply interpreted as an LTL formula. That is, 
if ASSUMPTIONS contains the LTL formula $\varphi_e$, INVARIANTS contains the 
formula $\psi_s$, and GUARANTEES contains the formula $\varphi_s$, then the 
specification is interpreted as 
\begin{equation*}
  \varphi_e \rightarrow \LTLglobally \psi_s \land \varphi_s
\end{equation*}
in standard LTL semantics (see Appendix~\ref{apx_ltl}).

\subsubsection{Strict semantics}
If the semantics is \mbox{\lstinline!Mealy,Strict!} or 
\mbox{\lstinline!Moore,Strict!}, and the TARGET coincides with the semantics 
system 
model, then the specification is interpreted under strict implication 
semantics (as used in the synthesis of GR(1) specifications). Essentially, 
this means that the system is only allowed to violate safety 
properties if the environment has violated the safety assumptions at an 
earlier time. This is in contrast to the standard LTL semantics, where the 
specification is satisfied if the environment violates the assumptions at any 
time, regardless of the system behavior. For details, see Klein and 
Pnueli~\cite{KleinP10} and Bloem et al.~\cite{DBLP:journals/jcss/BloemJPPS12}.

\subsection{Derived cases}

\subsubsection{Conversion between system models}
If the implementation model of the SEMANTICS differs from the TARGET of a 
specification, we use a simple conversion to get a specification that is 
realizable in the target system model iff the original specification is 
realizable in the original system model: a specification in 
Moore semantics can be converted into Mealy semantics by prefixing all 
occurrences of input atomic propositions with an additional 
$\LTLnext$-operator. Similarly, we can convert from Mealy semantics to Moore 
semantics by prefixing outputs with an $ \LTLnext $-operator.

\subsubsection{Conversion from strict to non-strict}
Finally, a specification in strict semantics can also be converted to an 
equivalent specification in standard semantics. Consider a specification with 
assumptions~$ \rho_{e} $, invariants~$ \psi_{s} $, and guarantees~$ \varphi_{s} $.
Assume that $\rho_e$ can be written as $\theta_e \land \LTLglobally \psi_e 
\land \varphi_e$, where $\theta_e$ is an \emph{initial constraint} (that 
only talks about the initial values of inputs), $\psi_e$ is a \emph{safety 
constraint}  (we assume that it is an invariant over the current and next 
value of inputs at any time), and $\varphi_e$ a \emph{liveness constraint}. 
Similarly, assume that $\psi_s$ is a safety constraint, and separate $\varphi_s$ 
into an initial constraint $\theta_s$ and a liveness constraint $\varphi_s'$. 
Then the formula~$ \theta_e \land \LTLglobally \psi_{e} \land \varphi_{e} 
\rightarrow
\theta_s \land \LTLglobally \psi_{s} \land \varphi_{s} $
under strict implication semantics can be converted to the following formula
in standard semantics\footnote{Note that in the conversion of
  \cite{DBLP:journals/jcss/BloemJPPS12}, the formula is strengthened by
  adding the
  formula~$ \LTLglobally (\LTLpastglobally \psi_{e} \rightarrow
  \psi_{s})$,
  where $ \LTLpastglobally \varphi $ is a Past-LTL formula and denotes
  that $ \varphi $ holds everywhere in the past. However, it is easy
  to show that our definition of strict semantics matches the
  definition of \cite{DBLP:journals/jcss/BloemJPPS12}. We prefer this
  notion, since it avoids the introduction of Past-LTL.}:
\begin{equation*}
  \theta_e \rightarrow \left( \theta_s \land (\psi_{s} \LTLweakuntil \neg \psi_{e}) \land (\LTLglobally \psi_{e} \land \varphi_{e} \rightarrow \LTLglobally \psi_{s} \wedge \varphi_{s}) \right).
\end{equation*}

\section{The Full Format}
\label{sec:format}

In the full format, a specification consists of three sections: the
INFO section, the GLOBAL section and the MAIN section. The GLOBAL
section is optional.
\begin{equation*}
  \tlsfsec{info} [\tlsfsec{global}] \tlsfsec{main}
\end{equation*}
The INFO section is the same as in the basic format, defined in
\secref{sec:basicinfo}. The GLOBAL section can be used to define
parameters, and to bind identifiers to expressions that can be used
later in the specification. The MAIN section is used as before, but
can use extended sets of declarations and expressions.

We define the GLOBAL section in \secref{sec:global}, and the changes
to the MAIN section compared to the basic format in
\secref{sec:main-full}. The extended set of expressions that can be
used in the full format is introduced in \secref{sec:expressions},
extended signal and function declarations in
Sections~\ref{sec:signals} and \ref{sec:functions}, and additional
notation in Sections~\ref{sec:bigoperator}--\ref{sec:comments}.

\subsection{The GLOBAL Section}
\label{sec:global}

The GLOBAL section consists of the PARAMETERS subsection, defining the
identifiers that parameterize the specification, and the DEFINITIONS
subsection, that allows to define functions and bind identifiers to
complex expressions. Multiple declarations need to be separated by a
'\lstinline!;!'. The section and its subsections are optional.

\vspace{1em}

\noindent
\lstinline!  GLOBAL {!\\%
\lstinline!    PARAMETERS { !\\%
\lstinline!      !$ ( \tlsfid{identifier} $\lstinline! = !%
$ \tlsfsec{numerical expression} $\lstinline!;!$ )^{*} $\\%
\lstinline!    }!\\%
\lstinline!    DEFINITIONS  { !\\%
\lstinline!      !%
$ ((\tlsfid{function declaration} \sep \tlsfid{identifier} $%
\lstinline! = !$ \tlsfsec{expression}) $\lstinline!;!$ )^{*} $\\%
\lstinline!    }!\\%
\lstinline!  }!

\subsection{The MAIN Section}
\label{sec:main-full}

Like in the basic format, the MAIN section contains the partitioning
of input and output signals, as well as the main
specification. However, signal declarations can now contain signal
buses, and LTL expressions can use parameters, functions, and
identifiers defined in the GLOBAL section.

\vspace{1em}

\noindent
\lstinline!  MAIN {!\\%
\lstinline!    INPUTS      { !%
$ ( \tlsfsec{signal declaration} $\lstinline!;!$ )^{*} $\lstinline! }!\\%
\lstinline!    OUTPUTS     { !%
$ ( \tlsfsec{signal declaration} $\lstinline!;!$ )^{*} $\lstinline! }!\\%
\lstinline!    ASSUMPTIONS { !%
$ ( \tlsfsec{LTL expression} $\lstinline!;!$ )^{*} $\lstinline! }!\\%
\lstinline!    INVARIANTS  { !%
$ ( \tlsfsec{LTL expression} $\lstinline!;!$ )^{*} $\lstinline! }!\\%
\lstinline!    GUARANTEES  { !%
$ ( \tlsfsec{LTL expression} $\lstinline!;!$ )^{*} $\lstinline! }!\\%
\lstinline!  }!

\vspace{1em}

\noindent
As before, the ASSUMPTIONS, INVARIANTS and GUARANTEES subsections are
optional.

\subsection{Expressions}
\label{sec:expressions}

An expression~$ e $ is either a boolean signal, an $ n $-ary signal
(called bus), a numerical expression, a boolean expression, an LTL
expression, or a set expression. Each expression has a corresponding
type that is either one of the basic types:
$ \signals, \busses, \nats, \bools, \temporals $, or a recursively
defined set type $ \mathcal{S}_{\arbitrary} $ for some type
$ \arbitrary $.

As before, an identifier is represented by a string consisting of
lowercase and uppercase
letters~(\mbox{'\lstinline!a!'-'\lstinline!z!'},
\mbox{'\lstinline!A!'-'\lstinline!Z!'}),
numbers~('\lstinline!0!'-'\lstinline!9!'),
underscores~('\lstinline!_!'), primes~('\lstinline!'!'), and
at-signs~('\lstinline!@!') and does not start with a number or a
prime.  In the full format, identifiers are bound to expressions of
different type. We denote the respective sets of identifiers by
$ \signalids $, $ \busids $, $ \natids $, $ \boolids $,
$ \temporalids $, and $ \atypeids $.  Finally, basic expressions can
be composed to larger expressions using operators. In the full format,
we do not require fully parenthesized expressions. If an expression is
not fully parenthesized, we use the precedence order given in
Appendix~\ref{apx_precedence}. An overview over the all types of
expressions and operators is given below.

\subsubsection{Numerical Expressions}
A numerical expression~$ e_{\nats} $ conforms to the following
grammar:
\begin{eqnarray*}
  e_{\nats} & \equiv &
  i \text{\ \ \ for } i \in \natids \sep 
  n  \text{\ \ \ for } n \in \nats \sep
  e_{\nats} \ \text{\lstinline!+!} \ e_{\nats} \sep
  e_{\nats} \ \text{\lstinline!-!} \ e_{\nats} \sep
  e_{\nats} \ \text{\lstinline!*!} \ e_{\nats} \sep
  e_{\nats} \ \text{\lstinline!/!} \ e_{\nats} \sep
  e_{\nats} \ \, \text{\lstinline!\%!} \ \, e_{\nats} \\ 
  & &
  \text{\lstinline!|!} e_{\mathcal{S}_{\arbitrary}} \text{\lstinline!|!} \sep
  \text{\lstinline!MIN!} \ e_{\mathcal{S}_{\nats}} \sep
  \text{\lstinline!MAX!} \ e_{\mathcal{S}_{\nats}} \sep
  \text{\lstinline!SIZEOF!} \ s \ \ \text{ for } s \in \busids
\end{eqnarray*}
Thus, a numerical expression either represents an identifier (bound to
a numerical value), a numerical constant, an addition, a subtraction, a
multiplication, an integer division, a modulo operation, the size of a
set, the minimal/maximal value of a set of naturals, or the size (i.e., width) of a
bus, respectively. The semantics are defined in the usual way.

\subsubsection{Set Expressions}
A set expression~$ e_{\mathcal{S}_{\arbitrary}} $, containing elements
of type $ \mathbb{X} $, conforms to the following grammar:
\begin{eqnarray*}
  e_{\mathcal{S}_{\arbitrary}} & \equiv &
  i  \text{\ \ \ for } i \in \atypeids \sep
  \text{\lstinline!\{!} \, e_{\arbitrary} \text{\lstinline!,!} \, e_{\arbitrary}
  \text{\lstinline!,!} \ldots \text{\lstinline!,!} \, e_{\arbitrary} \,
  \text{\lstinline!\}!} \sep
  \text{\lstinline!\{!} \, e_{\nats} \text{\lstinline!,!} \, e_{\nats} \, 
  \text{\lstinline!..!}\,  e_{\nats} \, \text{\lstinline!\}!} \sep 
  \\ &&
  e_{\mathcal{S}_{\arbitrary}} \, \text{\lstinline!(+)!} \ e_{\mathcal{S}_{\arbitrary}} \sep
  e_{\mathcal{S}_{\arbitrary}} \, \text{\lstinline!(*)!} \ e_{\mathcal{S}_{\arbitrary}} \sep
  e_{\mathcal{S}_{\arbitrary}} \, \text{\lstinline!(\\)!} \ e_{\mathcal{S}_{\arbitrary}}
\end{eqnarray*}
Thus, the expression~$ e_{\mathcal{S}_{\arbitrary}} $ either
represents an identifier (bound to a set of values of type
$ \arbitrary $), an explicit list of elements of type $ \arbitrary $,
a list of elements specified by a range (for $ \arbitrary = \nats $),
a union of two sets, an intersection or a difference,
respectively. The semantics of a range expression
\lstinline!{!$ x $\lstinline!,!$ y $\lstinline!..!$ z $\lstinline!}!
are defined for $ x < y $ via:
\begin{equation*}
  \{ n \in \nats \mid x \leq n \leq z \wedge \exists j.\ n = x +
  j \cdot (y-x) \}.
\end{equation*}
The semantics of all other expressions is defined as usual. Sets
contain either positive integers, boolean expressions, LTL
expressions, buses, signals, or other sets of a specific type.

\subsubsection{Boolean Expressions}
A boolean expression~$ e_{\bools} $ conforms to the following
grammar:
\begin{eqnarray*}
  e_{\bools} & \equiv & 
  i  \text{\ \ \ for } i \in \boolids \sep
  e_{\arbitrary} \ \text{\lstinline!IN!} \ e_{\mathcal{S}_{\arbitrary}} \sep
  \text{\lstinline|true|} \sep
  \text{\lstinline!false!} \sep
  \text{\lstinline|!|}\,e_{\bools} \sep
  \\ & & 
  e_{\bools} \ \text{\lstinline!&&!} \ e_{\bools} \sep
  e_{\bools} \ \text{\lstinline!||!} \ e_{\bools} \sep
  e_{\bools} \ \text{\lstinline!->!} \ e_{\bools} \sep
  e_{\bools} \ \text{\lstinline!<->!} \ e_{\bools} \sep
  \\ & & 
  e_{\nats} \ \text{\lstinline!==!} \ e_{\nats} \sep
  e_{\nats} \ \text{\lstinline~!=~} \ e_{\nats} \sep
  e_{\nats} \ \text{\lstinline!<!} \ e_{\nats} \sep
  e_{\nats} \ \text{\lstinline!<=!} \ e_{\nats} \sep
  e_{\nats} \ \text{\lstinline!>!} \ e_{\nats} \sep
  e_{\nats} \ \text{\lstinline!>=!} \ e_{\nats}
\end{eqnarray*}
Thus, a boolean expression either represents an identifier (bound to a
boolean value), a membership test, true, false, a negation, a
conjunction, a disjunction, an implication, an equivalence, or an
equation between two positive integers (equality, inequality, less
than, less or equal than, greater than, greater or equal than),
respectively. The semantics are defined in the usual way. Note that
signals are not allowed in a boolean expression, but only in an LTL
expression.

\subsubsection{LTL Expressions}
An LTL expression~$ \varphi $ conforms to the same grammar as a
boolean expression, except that it additionally includes signals and
temporal operators.
%
\begin{equation*}
  \varphi \ \, \equiv \ \, \ldots \sep
  i \text{\ \ for } i \in \temporalids \sep
  s \text{\ \ for } s \in \signalids \sep
  b \text{\lstinline![!} e_{\nats} 
  \text{\lstinline!]!} \text{ for } b 
  \in \busids \sep
  \text{\lstinline!X!} \ \varphi \sep
  \text{\lstinline!G!} \ \varphi \sep
  \text{\lstinline!F!} \ \varphi \sep
  \varphi \ \text{\lstinline!U!} \ \varphi \sep
  \varphi \ \text{\lstinline!R!} \ \varphi \sep
  \varphi \ \text{\lstinline!W!} \ \varphi
\end{equation*}
\goodbreak
\noindent
Thus, an LTL expression additionally can represent an identifier bound
to an LTL formula, a signal, an $e_{\nats} $-th signal of a bus, a
next operation, a globally operation, an eventually operation, an
until operation, a release operation, or a weak until operation,
respectively. Note that every boolean expression is also an LTL
expression, thus we allow the use of identifiers that are bound to
boolean expressions as well. A formal definition of the semantics of
the temporal operators can be found in Appendix~\ref{apx_ltl}.

\subsection{Signals and Buses}
\label{sec:signals}

A signal declaration consists of the name of the signal.  As for the
basic format, signals are declared as either input or output signals,
denoted by $ \inputs $ and $ \outputs $, respectively. A bus
declaration additionally specifies a signal width, i.e., a bus
represents a finite set of signals.
\begin{equation*}
  \langle \text{\textit{name}} \rangle \sep
  \langle \text{\textit{name}} \rangle \text{\lstinline![!} 
  e_{\nats} \text{\lstinline!]!}
\end{equation*}
In other words, a signal declaration \lstinline!s! specifies a signal
$ s \in \inputs \cup \outputs $, where a bus declaration
\lstinline!b[n]! specifies $ n $ signals~\lstinline!b[0]!,
\lstinline!b[1]!, $ \ldots $, \lstinline!b[n-1]!, with either
\lstinline!b[i]!$ \in \inputs $ for all $ i $, or
\lstinline!b[i]!$ \in \outputs $ for all $ i $.

Consider that we use \lstinline!b[i]! to access the $ i $-th value of
$ b $, i.e., we use the same syntax as for the declaration
itself\footnote{C-Array Syntax Style}. Also note that for the declared
signals~$ s $, we have
$ s \in \inputs \cup \outputs \subseteq \signalids $, and for the
declared buses $ b $, we have $ b \in \busids $.

\subsection{Function Declarations}
\label{sec:functions}

As another feature, one can declare (recursive) functions of arbitrary
arity inside the DEFINITIONS section. Functions can be used to define
simple macros, but also to generate complex formulas from a given set
of parameters. A declaration of a function of arity~$ n $ has the form
%
\begin{equation*}
  \tlsfid{function name} \text{\lstinline!(!}
  \tlsfid{arg\ensuremath{_{1}}} \text{\lstinline!,!} 
  \tlsfid{arg\ensuremath{_{2}}} \text{\lstinline!,!}  
  \ldots \text{\lstinline!,!}
  \tlsfid{arg\ensuremath{_{n}}} \text{\lstinline!) =\ !}
  (e_{c})^{+},
\end{equation*}
where
$ \tlsfid{arg\ensuremath{_{1}}}, \tlsfid{arg\ensuremath{_{2}}},
\ldots, \tlsfid{arg\ensuremath{_{n}}} $
are fresh identifiers that can only be used inside the
sub-expressions~$ e_{c} $. An expression~$ e_{c} $ conforms to the
following grammar:
\begin{equation*}
  e_{c} \ \, \equiv \ \, e \sep
  e_{\bools} \ \text{\lstinline!:!} \ e \sep
  e_{\mathbb{P}} \ \text{\lstinline!:!} \ e 
  \qquad \qquad \text{where } 
  \ \ e \ \, \equiv  \ \,
  e_{\nats} \sep
  e_{\bools} \sep
  e_{\mathcal{S}_{\arbitrary}} \sep
  \varphi
\end{equation*}
Thus, a function can be bound to any expression~$ e $, parameterized
in its arguments, which additionally may be guarded by some boolean
expression~$ e_{\bools} $, or a pattern match~$ e_{\pats} $. If the
regular expression~$ (e_{c})^{+} $ consists of more than one
expression~$ e_{c} $, then the function binds to the first expression
whose guard evaluates to \lstinline!true! (in the order of their
declaration). Furthermore, the special guard \lstinline!otherwise! can
be used, which evaluates to true if and only if all other guards
evaluate to \lstinline!false!. Expressions without a guard are
implicitly guarded by \lstinline!true!. All sub-expressions~$ e_{c} $
need to have the same type $ \arbitrary $. For every instantiation of
a function by given parameters, we view the resulting expression
$ e_{\arbitrary} $ as an identifier in~$ \Gamma_{\arbitrary} $, bound
to the result of the function application.
%
 
\subsubsection{Pattern Matching}
\label{sec_patterns}
Pattern matches are special guards of the form
\begin{equation*}
  e_{\pats} \equiv \ \, \varphi \ \lstilde \ \varphi' ,
  \vspace{-0.5em}
\end{equation*}
which can be used to describe different behavior depending on the
structure of an LTL expression. Hence, a guard~$ e_{\pats} $ evaluates
to \lstinline!true! if and only if $ \varphi $ and $ \varphi' $ are
structurally equivalent, with respect to their boolean and temporal
connectives. Furthermore, identifier names that are used in
$ \varphi' $ need to be fresh, since every identifier expression that
appears in $ \varphi' $ is bound to the equivalent sub-expression in
$ \varphi $, which is only visible inside the right-hand-side of the
guard. Furthermore, to improve readability, the special
identifier~\lstinline!_! (wildcard) can be used,
which always remains unbound. To clarify this feature, consider the
following function declaration:

\vspace{0.6em}

\noindent
\lstinline!  fun(f) =!\\%
\vspace{-0.4em}%
\lstinline!    f !$ \hspace{-1.5pt} \lstilde \hspace{-1.5pt} $%
\lstinline! a U _: a!\\%
\lstinline!    otherwise: X f!
   
\vspace{0.6em}

\noindent The function $ \textit{fun} $ gets an LTL formula $ f $ as a
parameter. If $ f $ is an until formula of the form
$ \varphi_{1} \LTLuntil \varphi_{2} $, then $ \textit{fun}(f) $ binds
to $ \varphi_{1} $, otherwise $ \textit{fun}(f) $ binds to
$ \LTLnext f $.

\subsection{Big Operator Notation}
\label{sec:bigoperator}

It is often useful to express parameterized expressions using ``big''
operators, e.g., we use $ \Sigma $ to denote a sum over multiple
sub-expressions, $ \Pi $ to denote a product, or $ \bigcup $ to denote
a union. It is also possible to use this kind of notion in this
specification format. The corresponding syntax looks as follows:
\begin{equation*}
  \tlsfid{op} \text{\lstinline![!} \, 
  \tlsfid{id\ensuremath{_{0}}} \, 
  \text{\lstinline!IN!} \; e_{\mathcal{S}_{\arbitrary_{0}}} \! 
  \text{\lstinline!,!}\, \tlsfid{id\ensuremath{_{1}}} \, 
  \text{\lstinline!IN!} \; e_{\mathcal{S}_{\arbitrary_{1}}} \! 
  \text{\lstinline!,!}\, \ldots \, 
  \text{\lstinline!,!} \,\tlsfid{id\ensuremath{_{n}}} \, 
  \text{\lstinline!IN!} \; e_{\mathcal{S}_{\arbitrary_{n}}} 
  \text{\lstinline!]!} \, e_{\arbitrary} 
\end{equation*}
Let $ x_{j} $ and $ S_{j} $ be the identifier represented by
$ \tlsfid{id\ensuremath{_{j}}} $ and the set represented by
$ e_{\mathcal{S}_{\arbitrary_{j}}}\! $, respectively. Further, let
$ \bigoplus $ be the mathematical operator corresponding to
$ \tlsfid{op} $. Then, the above expression corresponds to the
mathematical expression:
\begin{equation*}
  \bigoplus\limits_{x_{0} \in S_{0}} \ 
  \bigoplus\limits_{x_{1} \in S_{1}} \ \cdots \ 
  \bigoplus\limits_{x_{n} \in S_{n}}
  \big( e_{\arbitrary} )
\end{equation*}
Note that $ \tlsfid{id\ensuremath{_{0}}} $ is already bound in
expression $ e_{\mathcal{S}_{\arbitrary_{1}}} \!$,
$ \tlsfid{id\ensuremath{_{1}}} $ is bound in
$ e_{\mathcal{S}_{\arbitrary_{2}}} \!$, and so forth. The syntax is
supported by every operator
$ \tlsfid{op} \in \{ \text{\lstinline!+!},
\text{\lstinline!*!}, \text{\lstinline!(+)!},
\text{\lstinline!(*)!}, \text{\lstinline!&&!},
\text{\lstinline!||!} \} $.

\subsection{Syntactic Sugar}
\label{sec:syntacticsugar}

To improve readability, there is additional syntactic sugar, which can
be used beside the standard syntax. Let $ n $ and $ m $ be numerical
expressions, then

\begin{itemize}

\item \lstinline!X[!$ n $\lstinline!]!$ \; \varphi $ denotes a stack
  of $ n $ next operations, e.g.: \\[0.5em]
  \lstinline!  X[3] a!$ \ \, \equiv \ \, $ \lstinline!X X X a!

\item \lstinline!F[!$ n $\lstinline!:!$ m $\lstinline!]!$ \; \varphi $
  denotes that $ \varphi $ holds somewhere between the next $ n $ and
  $ m $ steps, e.g.: \\[0.5em]
  \lstinline!  F[2:3] a!$ \ \, \equiv \ \, $\lstinline!X X(a || X a)!

\item \lstinline!G[!$ n $\lstinline!:!$ m $\lstinline!]!$ \; \varphi $
  denotes that $ \varphi $ holds everywhere between the next $ n $ and
  $ m $ steps, e.g.: \\[0.5em]
  \lstinline!  G[1:3] a!$ \ \, \equiv \ \, $\lstinline!X(a && X(a && X a))!

\item $ \tlsfid{op} $\lstinline![!$ \, \ldots $\lstinline!,!$ \, n \, \circ_{1} 
  \tlsfid{id} \circ_{2} \, m \, $\lstinline!,!$ \ldots $\lstinline!]!$ \, e_{X} $
  denotes a big operator application, where  $ n \, \circ_{1} \tlsfid{id} 
  \circ_{2} \, m $ with $ \circ_{1}, \circ_{2} \in \{ \text{\lstinline!<!},
  \text{\lstinline!<=!} \} $ denotes that \tlsfid{id} ranges from $ n $ to
  $ m $. Thereby, the inclusion of $ n $ and $ m $ depends on the
  choice of $ \circ_{1} $ and $ \circ_{2} $, respectively. Thus, the
  notation provides an alternative to membership in combination with
  set ranges, e.g.: \\[0.5em]
  \lstinline!  &&[0 <= i < n] b[i]!$ \ \, \equiv \ \, 
  $\lstinline!&&[i IN {0,1..n-1}] b[i]!
  %

\end{itemize}

\subsection{Comments}
\label{sec:comments}

It is possible to use C style comments anywhere in the specification,
i.e., there are single line comments initialized by \lstinline!//! and
multi line comments between \lstinline!/*! and
\lstinline!*/!.  Multi line comments can be nested.

\newpage

\section{Example: A Parameterized Arbiter}
\label{sec:example}

To get some feeling for the interplay of the aforementioned features,
consider the following example specification of a parameterized
arbiter.

\vspace{1em}

\noindent
\lstinline!  INFO!%
\lstinline[basicstyle=\ttfamily\color{black}]! {!\\%
\lstinline!    TITLE!%
\lstinline[basicstyle=\ttfamily\color{black}]!:       !%
\lstinline[stringstyle=\ttfamily\color{red!30!orange!80!black}]!"A Parameterized Arbiter"!\\%
\lstinline!    DESCRIPTION!%
\lstinline[basicstyle=\ttfamily\color{black}]!: !%
\lstinline[stringstyle=\ttfamily\color{red!30!orange!80!black}]!"An arbiter, parameterized in the number of clients"!\\%
\lstinline!    SEMANTICS!%
\lstinline[basicstyle=\ttfamily\color{black}]!:   !%
\lstinline[identifierstyle=\ttfamily\color{gray!70!black}]!Mealy!\\%
\lstinline!    TARGET!%
\lstinline[basicstyle=\ttfamily\color{black}]!:      !%
\lstinline[identifierstyle=\ttfamily\color{gray!70!black}]!Mealy!\\%
\lstinline[basicstyle=\ttfamily\color{black}]!  }!\\%
\lstinline!  !\\%
\lstinline!  GLOBAL !%
\lstinline[basicstyle=\ttfamily\color{black}]!{!\\%
\lstinline!    PARAMETERS !%
\lstinline[basicstyle=\ttfamily\color{black}]!{!\\%
\lstinline[commentstyle=\ttfamily\color{orange!60!black}]!      // two clients!\\%
\lstinline[identifierstyle=\ttfamily\color{green!50!black}]!      n!%
\lstinline[basicstyle=\ttfamily\color{black}]! = 2;!\\%
\lstinline[basicstyle=\ttfamily\color{black}]!    }!\\%
\lstinline!    DEFINITIONS!%
\lstinline[basicstyle=\ttfamily\color{black}]! {!\\%
\lstinline[commentstyle=\ttfamily\color{orange!60!black}]!      // mutual exclusion!\\%
\lstinline[identifierstyle=\ttfamily\color{green!50!black}]!      mutual!%
\lstinline[identifierstyle=\ttfamily\color{black},basicstyle=\ttfamily\color{black}]!(b) =!\\%
\lstinline[basicstyle=\ttfamily\color{cyan!50!black}]!        ||!%
\lstinline[identifierstyle=\ttfamily\color{black},basicstyle=\ttfamily\color{black}]![i !%
\lstinline[identifierstyle=\ttfamily\color{cyan!50!black}]!IN!%
\lstinline[basicstyle=\ttfamily\color{black},identifierstyle=\ttfamily\color{black}]! {0,1..n-1}]!\\%
\lstinline[basicstyle=\ttfamily\color{cyan!50!black}]!          &&!%
\lstinline[identifierstyle=\ttfamily\color{black},basicstyle=\ttfamily\color{black}]![j !%
\lstinline[identifierstyle=\ttfamily\color{cyan!50!black}]!IN!%
\lstinline[basicstyle=\ttfamily\color{black},identifierstyle=\ttfamily\color{black}]! {0,1..n-1} !%
\lstinline[basicstyle=\ttfamily\color{cyan!50!black}]!(\)!%
\lstinline[basicstyle=\ttfamily\color{black},identifierstyle=\ttfamily\color{black}]! {i}]!\\%
\lstinline[basicstyle=\ttfamily\color{cyan!50!black}]~            !~%
\lstinline[basicstyle=\ttfamily\color{black},identifierstyle=\ttfamily\color{black}]!(b[i] !%
\lstinline[basicstyle=\ttfamily\color{cyan!50!black}]!&&!%
\lstinline[basicstyle=\ttfamily\color{black},identifierstyle=\ttfamily\color{black}]! b[j]);!\\%
\lstinline[commentstyle=\ttfamily\color{orange!60!black}]!      // the Request-Response condition!\\%
\lstinline[identifierstyle=\ttfamily\color{green!50!black}]!      reqres!%
\lstinline[basicstyle=\ttfamily\color{black},identifierstyle=\ttfamily\color{black}]!(req,res) =!\\%
\lstinline[identifierstyle=\ttfamily\color{cyan!50!black}]!        G!%
\lstinline[basicstyle=\ttfamily\color{black},identifierstyle=\ttfamily\color{black}]! (req !%
\lstinline[basicstyle=\ttfamily\color{cyan!50!black},identifierstyle=\ttfamily\color{cyan!50!black}]!-> F!%
\lstinline[basicstyle=\ttfamily\color{black},identifierstyle=\ttfamily\color{black}]! res);!\\%
\lstinline[basicstyle=\ttfamily\color{black}]!    }!\\%
\lstinline[basicstyle=\ttfamily\color{black}]!  }!\\%
\lstinline!  !\\%
\lstinline[commentstyle=\ttfamily\color{orange!60!black}]!  /* Ensure mutual exclusion on the output bus and guarantee!\\%
\lstinline[basicstyle=\ttfamily\color{orange!60!black},identifierstyle=\ttfamily\color{orange!60!black}]!     that each request on the input bus is eventually granted */!\\%
\lstinline!  MAIN!%
\lstinline[basicstyle=\ttfamily\color{black}]! {!\\%
\lstinline!    INPUTS!%
\lstinline[basicstyle=\ttfamily\color{black}]! {!\\%
\lstinline[identifierstyle=\ttfamily\color{green!50!black}]!      r!%
\lstinline[basicstyle=\ttfamily\color{black},identifierstyle=\ttfamily\color{black}]![n];!\\%
\lstinline[basicstyle=\ttfamily\color{black}]!    }!\\%
\lstinline!    OUTPUTS!%
\lstinline[basicstyle=\ttfamily\color{black}]! {!\\%
\lstinline[identifierstyle=\ttfamily\color{green!50!black}]!      g!%
\lstinline[basicstyle=\ttfamily\color{black},identifierstyle=\ttfamily\color{black}]![n];!\\%
\lstinline[basicstyle=\ttfamily\color{black}]!    }!\\%
\lstinline!    INVARIANTS!%
\lstinline[basicstyle=\ttfamily\color{black}]! {!\\%
\lstinline[basicstyle=\ttfamily\color{black},identifierstyle=\ttfamily\color{black}]!      mutual(g);!\\%
\lstinline[basicstyle=\ttfamily\color{black}]!    }!\\%
\lstinline!    GUARANTEES!%
\lstinline[basicstyle=\ttfamily\color{black}]! {!\\%
\lstinline[basicstyle=\ttfamily\color{cyan!50!black}]!      &&!%
\lstinline[basicstyle=\ttfamily\color{black},identifierstyle=\ttfamily\color{black}]![0 <= i < n]!\\%
\lstinline[basicstyle=\ttfamily\color{black},identifierstyle=\ttfamily\color{black}]!        reqres(r[i],g[i]);!\\%
\lstinline[basicstyle=\ttfamily\color{black}]!    }!\\
\lstinline[basicstyle=\ttfamily\color{black}]!  }!

\vspace{1em}

\noindent The example is parameterized in the number of clients~$ n $
(here: $ n = 2 $). Furthermore, it uses two functions:
$ \textit{mutual}(b) $, which ensures mutual exclusions on the signals
of a bus~$ b $ of width~$ n $, and
$ \textit{reqres}(\textit{req}, \textit{res}) $, which ensures that
every request~$ \textit{req} $ is eventually followed by some
response~$ \textit{res} $. In the final specification, both conditions
are then combined over the inputs $ r_{i} $ and outputs $ g_{i} $.

\section{The SyFCo Tool}
\label{sec:tool}

We created a Synthesis Format Conversion Tool (SyFCo)~\cite{SyFCo}
that can interpret the high level constructs of the format and
supports transformation of the specification back to plain LTL. The
tool has been designed to be modular with respect to the supported
output formats and semantics. Furthermore, the tool can identify and
manipulate parameters, targets and semantics of a specification on the
fly, and thus allows comparative studies, as it is for example needed
in the Synthesis Competition.

\vspace{1em}

\noindent The main features of the tool can be summarized as follows:

\begin{itemize}

\item Evaluation of high level constructs in the full format to reduce
  full TLSF to basic TLSF.

\item Transformation to other existing specification formats, like
  Promela LTL~\cite{PromelaLTL}, PSL~\cite{EF2006},
  Unbeast~\cite{E2010}, or Wring~\cite{SomenziB00}.

\item On the fly adjustment of parameters, semantics or targets.

\item Preprocessing of the resulting LTL formula

\begin{itemize}

\item[$ \circ $] conversion to negation normal form

\item[$ \circ $] replacement of derived operators

\item[$ \circ $] pushing/pulling next, eventually, or globally operators inwards/outwards

\item[$ \circ $] $ \ldots $ \raisebox{-8pt}{\ }

\end{itemize}

\end{itemize}

\section{Extensions}
\label{sec:extensions}

The format remains open for further extensions, which allow more fine
grained control over the specification with respect to a particular
synthesis problem. At the current time of writing, the following
extensions were under consideration:

\begin{itemize}

\item Compositionality: The possibility to separate specifications
  into multiple components, which then can be used as building blocks
  to specify larger components.
	
\item Partial Implementations: a specification that is separated into
  multiple components might also contain components that are already
  implemented. Implemented components could be given in the AIGER
  format that is already used in SYNTCOMP~\cite{SYNTCOMP-format}.

\item Additional syntactic sugar, like enumerations or arithmetic on
  busses.

\end{itemize}

\section*{Acknowledgments}

We thank Sebastian Schirmer, who supported us with the results of his
Bachelor Thesis~\cite{THSEBSCH}, to resolve many design decisions
that came up during the development of this format. We thank Roderick Bloem, 
R\"udiger Ehlers, Bernd Finkbeiner, Ayrat Khalimov, Robert K\"onighofer, Nir 
Piterman, and Leander Tentrup for comments on the TSLF and drafts of this 
document.

\bibliographystyle{plain}
\bibliography{biblio}

\newpage
\appendix

\section{Appendix}

\subsection{Linear Temporal Logic}
\label{apx_ltl}

Linear Temporal Logic (LTL) is a temporal logic, defined over a finite
set of atomic propositions~$ \text{AP} $. The syntax of LTL conforms to the
following grammar:
\begin{equation*}
  \varphi \ \ := \ \ \text{\textit{true}} \sep  p \in \text{AP}
  \sep \neg \varphi \sep \varphi \vee \varphi
  \sep \LTLnext \varphi \sep \varphi \LTLuntil \varphi
\end{equation*}
The semantics of LTL are defined over infinite
words~$ \alpha = \alpha_{0}\alpha_{1}\alpha_{2}\dots \in
(2^{\text{AP}})^{\omega} $.
A word~$ \alpha $ satisfies a formula~$ \varphi $ at position
$ i \in \nats $:

\begin{itemize}

\item $ \alpha, i \vDash \textit{true} $ 

\item $ \alpha, i \vDash p $ \ iff \ $ p \in \alpha_{i} $

\item $ \alpha, i \vDash \neg \varphi $ \ iff \
  $ \alpha, i \not\vDash \varphi $

\item $ \alpha, i \vDash \varphi_{1} \vee \varphi_{2} $ \ iff \
  $ \alpha, i \vDash \varphi_{1} $ or $ \alpha, i \vDash \varphi_{2} $

\item $ \alpha, i \vDash \LTLnext \varphi $ \ iff \
  $ \alpha, i + 1 \vDash \varphi $

\item $ \alpha, i \vDash \varphi_{1} \LTLuntil \varphi_{2} $ \ iff \
  $ \exists n \geq i.\ \alpha, n \vDash \varphi_{2} $ and
  $ \forall i \leq j < n.\ \alpha, j \vDash \varphi_{1} $

\end{itemize}

\noindent A word~$ \alpha \in 2^{\text{AP}} $
satisfies a formula~$ \varphi $ iff $ \alpha, 0 \vDash \varphi $.
Beside the standard operators, we have the following derived operators:

\begin{itemize}

\item
  $ \varphi_{1} \wedge \varphi_{2} \equiv \neg(\neg \varphi_{1} \vee
  \neg \varphi_{2}) $

\item
  $ \varphi_{1} \rightarrow \varphi_{2} \equiv \neg \varphi_{1} \vee
  \varphi_{2} $

\item
  $ \varphi_{1} \leftrightarrow \varphi_{2} \equiv (\varphi_{1}
  \rightarrow \varphi_{2}) \wedge (\varphi_{2} \rightarrow
  \varphi_{1}) $

\item
  $ \LTLeventually \varphi \equiv \text{\textit{true}} \LTLuntil
  \varphi $

\item $ \LTLglobally \varphi \equiv \neg \LTLeventually \neg \varphi $

\item
  $ \varphi_{1} \LTLrelease \varphi_{2} \equiv \neg(\neg \varphi_{1}
  \LTLuntil \neg \varphi_{2}) $

\item
  $ \varphi_{1} \LTLweakuntil \varphi_{2} \equiv (\varphi_{1}
  \LTLuntil \varphi_{2}) \vee \LTLglobally \varphi_{1} $

\end{itemize}

\subsection{Mealy and Moore Automata}
\label{apx_memo}

A Mealy automaton is a tuple
$ \mathcal{M}_{e} = (\inputs, \outputs, Q, q_{0}, \delta, \lambda_{e})
$, where

\begin{itemize}

\item $ \inputs $ is a finite set of input letters,

\item $ \outputs $ is a finite set of output letters,

\item $ Q $ is finite set of states,

\item $ q_{0} \in Q $ is the initial state,

\item $ \delta \colon Q \times \inputs \rightarrow Q $ is the
  transition function, and

\item
  $ \lambda_{e} \colon Q \times \inputs \rightarrow \outputs $
  is the output function.

\end{itemize}

\noindent Hence, the output depends on the current state of the
automaton and the last input letter.

\bigskip

\noindent A Moore automaton is a tuple
$ \mathcal{M}_{o} = (\inputs, \outputs, Q, q_{0}, \delta, \lambda_{o})
$,
where $ \inputs, \outputs, Q, q_{0} $ and $ \delta $ are defined as
for Mealy automata. However, the output function
$ \lambda_{o} \colon Q \rightarrow \outputs $ determines the current
output only on the current state of the automaton, but not the last
input.

\subsection{Operator Precedence, Alternative Operators}
\label{apx_precedence}

The following table lists the precedence, arity and associativity of
all expression operators. Also consider the alternative names in
brackets which can be used instead of the symbolic representations.

\renewcommand{\arraystretch}{1.1}

\begin{center}
  \centering

  \begin{tabular}{|c|l|l|c|c|}
    \hline 
    \textbf{Precedence} & \textbf{Operator} & \textbf{Description} & \textbf{Arity} &  \textbf{Associativity} \\
    \hline
    \hline
    \multirow{6}{*}{1} & \lstinline!+[!$ \cdot $\lstinline!]!\ (\lstinline!SUM[!$ \cdot $\lstinline!]!) & sum & \multirow{6}{*}{unary} & \\
      & \lstinline!*[!$ \cdot $\lstinline!]!\ (\lstinline!PROD[!$ \cdot $\lstinline!]!) & product & & \\
      & \lstinline!|!$ \cdots $\lstinline!|!\ (\lstinline!SIZE!) & size & & \\
      & \lstinline!MIN! & minimum & & \\
      & \lstinline!MAX! & maximum & & \\
      & \lstinline!SIZEOF! & size of a bus & & \\
    \hline
    2 & \lstinline!*!\ (\lstinline!MUL!) & multiplication & binary & left-to-right \\
    \hline
    \multirow{2}{*}{3} & \lstinline!/!\ (\lstinline!DIV!) & integer division & \multirow{2}{*}{binary} & \multirow{2}{*}{right-to-left} \\
      & \lstinline!%!\ (\lstinline!MOD!) & modulo & & \\
    \hline
    \multirow{2}{*}{4} & \lstinline!+!\ (\lstinline!PLUS!) & addition & \multirow{2}{*}{binary} & \multirow{2}{*}{left-to-right} \\
      & \lstinline!-!\ (\lstinline!MINUS!) & difference & & \\
    \hline 
    \multirow{2}{*}{5} & \lstinline!(*)[!$ \cdot $\lstinline!]!\ (\lstinline!CAP[!$ \cdot $\lstinline!]!) & intersection & \multirow{2}{*}{unary} & \\
      & \lstinline!(+)[!$ \cdot $\lstinline!]!\ (\lstinline!CUP[!$ \cdot $\lstinline!]!) & union &  & \\
    \hline
    6 & \lstinline!(\)!\ (\lstinline!(-)!,\lstinline!SETMINUS!) & set difference & binary & right-to-left \\
    \hline
    7 & \lstinline!(*)!\ (\lstinline!CAP!) & intersection & binary & left-to-right \\
    \hline
    8 & \lstinline!(+)! (\lstinline!CUP!) & union & binary & left-to-right \\
    \hline
    \multirow{6}{*}{9} & \lstinline!==!\ (\lstinline!EQ!) & equality & \multirow{6}{*}{binary} & \multirow{6}{*}{left-to-right} \\
      & \lstinline~!=~\ (\lstinline!/=!, \lstinline!NEQ!) & inequality & & \\
      & \lstinline!<!\ (\lstinline!LE!) & smaller than & & \\
      & \lstinline!<=!\ (\lstinline!LEQ!) & smaller or equal than & & \\
      & \lstinline!>!\ (\lstinline!GE!) & greater then & & \\
      & \lstinline!>=!\ (\lstinline!GEG!) & greater or equal than & & \\
    \hline
    10 & \lstinline!IN!\ (\lstinline!ELEM!, \lstinline!<-!) & membership & binary & left-to-right \\
    \hline
    \multirow{6}{*}{11} & \lstinline~!~\ (\lstinline!NOT!) & negation & \multirow{6}{*}{unary} & \\
       & \lstinline!X! & next & & \\
       & \lstinline!F! & finally & & \\
       & \lstinline!G! & globally & & \\
       & \lstinline!&&[!$ \cdot $\lstinline!]!\ (\lstinline!AND[!$ \cdot $\lstinline!]!, \lstinline!FORALL[!$ \cdot $\verb!]!) & conjunction & & \\
       & \lstinline!||[!$ \cdot $\lstinline!]!\ (\lstinline!OR[!$ \cdot $\lstinline!]!, \lstinline!EXISTS[!$ \cdot $\verb!]!) & disjunction & & \\
    \hline
    12 & \lstinline!&&!\ (\lstinline!AND!) & conjunction & binary & left-to-right \\
    \hline
    13 & \lstinline!||!\ (\lstinline!OR!) & disjunction & binary & left-to-right \\     
    \hline
    \multirow{2}{*}{14} & \lstinline!->!\ (\lstinline!IMPLIES!) & implication & \multirow{2}{*}{binary} & \multirow{2}{*}{right-to-left} \\   
       & \lstinline!<->!\ (\lstinline!EQUIV!) & equivalence &  & \\     
    \hline
    15 & \lstinline!W! & weak until & binary & right-to-left \\
    \hline
    16 & \lstinline!U! & until & binary & right-to-left \\
    \hline
    17 & \lstinline!R! & release & binary & left-to-right \\
    \hline
    18 & \!{\color{blue!70!black}\raisebox{-9pt}{\scalebox{1.7}{\textasciitilde}}} \vspace{-5pt} & pattern match & binary & left-to-right \\
    \hline
    19 & \lstinline!:! & guard & binary & left-to-right \\    
    \hline
  \end{tabular}
\end{center}

\end{document}